\documentclass[11pt]{article}
\usepackage[margin=1in]{geometry}
\usepackage{graphicx}
\usepackage{booktabs}
\usepackage{amsmath}
\usepackage{amssymb}
\usepackage[hidelinks]{hyperref}
\usepackage{xcolor}

\title{Factions Within, Uncertain Across:\\
Within-Document Reader Sub-Groups in Social Highlighting}
\author{
  Kazuki Nakayashiki \quad Keisuke Watanabe \\[3pt]
  Glasp Inc. \\
  \texttt{kazuki@glasp.co} \quad \texttt{kei@glasp.co} \\[3pt]
  {\small\itshape Co-first authors (equal contribution).}
}
\date{}

\begin{document}
\maketitle

\begin{abstract}
When many people highlight the same document, is the ``crowd'' a single consensus, or is it
internally structured into reader sub-groups that mark different things --- and if so, is that
structure a stable property of a reader or of the document? This matters for personalization:
stable reader sub-populations could be learned from history and used to produce segmented
popular highlights for documents no one has read yet. Building on prior work showing that an
\emph{individual's} within-document signal is a whisper while individuality lives in
\emph{selection}~\cite{nakayashiki2026salience, nakayashiki2026selection}, we ask the
group-level question on a co-readership platform, using a margin-preserving \emph{curveball}
null. \textbf{Experiment~1:} within a document, readers form strong sub-groups --- pairs agree
with each other far beyond what shared salience, mark density, and sentence popularity predict
(nearest-neighbour agreement $z=+6.3$, significant in $88\%$ of documents; the standardized
effect is density-dependent, so we also report it in agreement units). Under an eight-block
region-preserving null, shared engagement with the same coarse regions of the document accounts
for about $40\%$ of this excess; the majority survives as finer reader-specific agreement
($z=+3.6$, $77\%$ significant). So the within-document crowd is, in this descriptive sense, factional rather than
a single consensus. \textbf{Experiment~2:} is that grouping a stable reader trait? Here we are honest
about power. The cross-document split-half reproducibility of a pair's agreement is near zero
pooled (e.g.\ $+0.078$ and $0.000$ in two separately drawn samples), and a power calibration shows the
test is informative only for pairs that co-read \emph{many} documents. In the only informative
high-overlap subset ($k\geq 4$), point estimates are positive but small-sample, imprecise across
the separately drawn samples, never significant, and attenuate under the region-preserving null.
\emph{No stratum is significant.} We therefore leave cross-document stability \textbf{unresolved}:
the data is consistent with anything from situational grouping to a weak-to-moderate stable reader
trait, and current co-readership density cannot separate them. The crowd is factional within a
document; whether its factions follow the reader across documents is, honestly, beyond our
reach.
\end{abstract}

\section{Introduction}
Aggregated highlights --- the passages many readers mark in the same document --- behave as a
collective signal of what matters, and platforms surface them as ``popular highlights.'' A
natural reading is that this crowd is a single consensus: a ranking of sentences by how many
readers marked them. But a crowd can be a \emph{mixture}. The people who co-read an article
about, say, a new model might split into those who mark the benchmark numbers and those who
mark the method; a consensus ranking averages them into one map and hides the split.

Whether that split exists, and whether it is a stable property of \emph{readers} or of
\emph{documents}, is the subject of this paper. It is the group-level companion to a finding
from our prior work: \emph{which sentences} an individual highlights within a document is mostly
shared salience --- a person's own history predicts their marks barely better than another
reader's does (own-versus-other gap $+0.017$ average precision, a
whisper)~\cite{nakayashiki2026salience} --- while the recoverable individual signal lives in
\emph{selection} (which already-salient things are theirs), where it is real but modest and
largely thematic~\cite{nakayashiki2026salience, nakayashiki2026selection}. The individual,
within a document, is nearly invisible. The open possibility is that \emph{groups} are not:
individual idiosyncrasy may be a whisper, yet align into sub-groups that, in aggregate, mark
systematically different things.

We ask two questions. \textbf{RQ1}: on a single co-read document, do readers form sub-groups
that agree with each other more than chance --- beyond shared salience, and beyond merely
engaging the same regions of the document? \textbf{RQ2}: is any such sub-grouping a
\emph{stable} reader trait --- do the same readers keep grouping together across the different
documents they co-read, the way a genuine sub-population (novices, experts) would? RQ2 is
decisive for personalization: a stable trait could be learned from a reader's history and used
to segment a cold document; a document-specific grouping could not.

Our answers are asymmetric in confidence. RQ1 is a clear \emph{yes}, and we decompose it:
within-document sub-structure is strong, about $40\%$ of it is shared region engagement, and
the majority is a finer reader-specific component that survives region-matching. RQ2 we cannot answer: the test is
underpowered at the available co-readership density, the informative high-overlap subset is
positive but imprecise and never significant --- so neither the existence nor the size of a stable
component can be established --- and a single pooled estimate misleads. The honest result is that the within-document crowd is factional, but the portability
of its factions is unresolved.

\paragraph{Contributions.}
(1) A group-level extension of the salience/selection picture: within-document highlighting
contains \emph{strong} reader sub-group structure, measured against a margin-preserving null
that removes shared salience, density, and popularity --- the loud counterpart of the
individual whisper. (2) A decomposition of that structure into shared \emph{region engagement}
($\sim$40\% of the raw excess) and a larger finer reader-specific residual ($\sim$60\%), via a
region-preserving null. (3) An honest,
power-calibrated, $k$-stratified treatment of cross-document stability that leaves the question
open rather than overclaiming, and surfaces a pooling pitfall: a single pooled estimate cannot
distinguish a confident zero from a zero produced by no power, reading as a settled null when the
question is actually undecided. (4) A reusable methodology: curveball and region-preserving
permutation nulls for reader-agreement structure, and synthetic calibration of what the
stability test can and cannot detect.

All analyses use margin-preserving nulls and are validated on synthetic ground truth; the main
effect sizes carry cluster-bootstrap confidence intervals.\footnote{Paper and figures are released at \url{https://github.com/glasp-co/clone-crowd}.
The scoring pipeline runs against private user highlighting behaviour and is not released;
per-pair results derive from individual behaviour and are not published.}

\section{Related work}
\paragraph{Highlighting as shared, but not necessarily monolithic, salience.}
Aggregated highlights act as collective reader response~\cite{winchell2020}, and our prior work
quantified the split between shared salience and the individual
residual~\cite{nakayashiki2026salience, nakayashiki2026selection}; there the own-versus-other
within-document gap was $+0.017$ (a whisper) and selection-level individuality was $\sim$$+0.14$,
mostly thematic. Those papers measured the \emph{individual}; here we ask whether the shared
layer is itself structured into groups.

\paragraph{Disagreement is signal; crowds are mixtures.}
Perspectivist and crowd-truth work argues that annotator disagreement is signal, not noise, and
that larger, more diverse pools surface lower agreement~\cite{aroyo2015, plank2022}. RQ1 is a
naturalistic instance: we ask whether the disagreement within a document's readers is
\emph{structured} (sub-groups) rather than uniform. The wisdom-of-crowds literature explains why
aggregates are strong --- diverse errors cancel~\cite{surowiecki2004} --- but a strong aggregate
can still hide sub-group structure, which a consensus ranking averages away.

\paragraph{Sub-populations, personas, and silicon crowds.}
A central motivation for modelling sub-populations is generative: ensembles of diverse models
can rival human crowds at forecasting~\cite{schoenegger2024}, and generative agents simulate
populations~\cite{park2024}. But prompt-built personas of a single model tend to collapse to a
mode, so genuine diversity must come from real behavioural grounding. Our results speak to the
precondition for such aggregation in reading: behavioural sub-structure exists within documents,
but whether it is stable enough to condition a per-reader model on remains open.

\paragraph{Personalized highlighting.}
Personalized highlight and summary models condition on a user's history~\cite{gygli2018,
salemi2023}. Our prior work found this does not pay off at the sentence level because salience
is shared~\cite{nakayashiki2026selection}; the present paper asks whether \emph{group}
membership could, and finds the stability such a method would need is unconfirmed at the current
co-readership density.

\section{Data and method}
Glasp is a social web highlighter; for a document marked by multiple users we know exactly who
highlighted what, materializing a co-readership structure. We use public-web articles only
(excluding Kindle and PDF). We reuse the within-document object of~\cite{nakayashiki2026salience}:
fetch the article body, segment it into sentences, and re-anchor each reader's highlighted
spans to sentences (a drift gate keeps a reader only if at least half of their spans re-anchor),
giving, per document, a \textbf{reader\,$\times$\,sentence binary matrix}. The analysis uses
only these binary marks --- \emph{no embeddings and no language model} --- so the structure we
measure is agreement on \emph{which sentences}, not semantic similarity.

\paragraph{Agreement beyond chance: margin-preserving nulls.}
Two readers overlap partly because everyone marks the introduction. To isolate sub-group
structure from this, we compare observed agreement to a \textbf{curveball} permutation null: we
randomly permute the binary matrix while preserving \emph{both} margins --- each reader's mark
count (rows) and each sentence's popularity (columns) --- so only \emph{which reader holds which
marks} is randomized. Agreement that survives this null is residual reader-agreement structure
beyond shared salience, density, and popularity (which we interpret, descriptively, as reader
sub-grouping --- without claiming validated discrete communities; see \S\ref{sec:e1}). Pairwise agreement is the binary cosine $\lvert A\cap B\rvert /
\sqrt{\lvert A\rvert\,\lvert B\rvert}$ between two readers' sentence sets.

A critic could still object that readers highlight \emph{contiguously} --- the same paragraph,
section, intro, or conclusion --- so the curveball signal might reflect readers engaging the
same \emph{regions} rather than a portable reader type. We therefore add a stricter
\textbf{region-preserving null}: a curveball applied independently within position blocks, which
preserves each reader's mark count \emph{per region} as well as both margins, randomizing only
which sentence \emph{within} a region they mark (we approximate regions by eight equal
sentence-position blocks; true section boundaries would be a sharper control, and because
position blocks only partly capture sections, the surviving signal is an \emph{upper} estimate
of the finer component). Structure that survives this null is finer than shared region
engagement.

\paragraph{The two tests.}
\textbf{RQ1 (existence; Experiment~1).} Per document we compute the mean \emph{nearest-neighbour
agreement} (does every reader have a close partner?) and the \emph{variance} of pairwise
agreement, and a per-document $z$-score and permutation $p$-value against each null, then
aggregate over documents. \textbf{RQ2 (stability; Experiment~2).} For reader pairs who co-read
$k\geq 2$ documents we compute, per shared document, the curveball-\emph{excess} agreement
$e_{AB}(D)=\text{obs cosine}-\text{curveball null mean}$ (unbiased at any reader count), and test
whether a pair's excess is a reproducible trait via the \textbf{split-half correlation} of
per-pair mean excess across the documents the pair co-reads. A stable sub-population makes a
pair's agreement consistent across documents (positive split-half); a document-specific grouping
does not.

\paragraph{Validation first.}
Before trusting either test on real data we calibrate it on synthetic communities with known
ground truth: a \emph{no-structure} control (readers independent given the marginal), a
\emph{planted-group} control (two fixed sub-groups), and --- for RQ2 --- a \emph{stable} control
(group membership persists across documents) versus a \emph{situational} control (membership
reshuffled per document), at several data densities. This is how we learn what each test can and
cannot detect.

\paragraph{Sampling and fetch.}
We sample co-readership communities (seed documents, their co-readers, and those readers' other
highlighted documents) and keep documents dense enough for a stable null. Live fetching of
arbitrary highlighted URLs fails often (paywalls, JavaScript, dead links); a content-extraction
service roughly doubled the usable yield. Fetch failures reduce sample size and may limit
external validity (the surviving documents skew toward easily-fetched public articles), though
they do not mechanically induce agreement beyond the document-specific null. All intervals are
cluster bootstraps (by document for RQ1; by pair, and separately by reader, for RQ2).
Table~\ref{tab:data} summarizes the samples; further detail is in Appendix~\ref{app:repro}.

\begin{table}[t]
\centering
\small
\begin{tabular}{lcc}
\toprule
 & Exp.~1 (structure) & Exp.~2 (stability) \\
\midrule
community seed documents       & 300 & 200 \\
dense documents                & 293 ($\geq 10$ readers) & 407 ($\geq 6$ readers) \\
usable after fetch + anchoring  & 75 & 146 \\
median readers / document       & 25 & --- \\
median sentences / document     & 74 & --- \\
reader pairs, cross-document ($k\geq 2$) & --- & 744 \\
distinct readers in those pairs & --- & 252 \\
\bottomrule
\end{tabular}
\caption{Samples. ``Dense'' documents have enough co-readers for a stable null; the
fetch+anchoring yield ($\sim$25--35\% of dense targets) is the main sample-size constraint.}
\label{tab:data}
\end{table}

\section{Experiment 1: reader sub-group structure within a document}
\label{sec:e1}
\begin{figure}[t]
\centering
\includegraphics[width=0.84\linewidth]{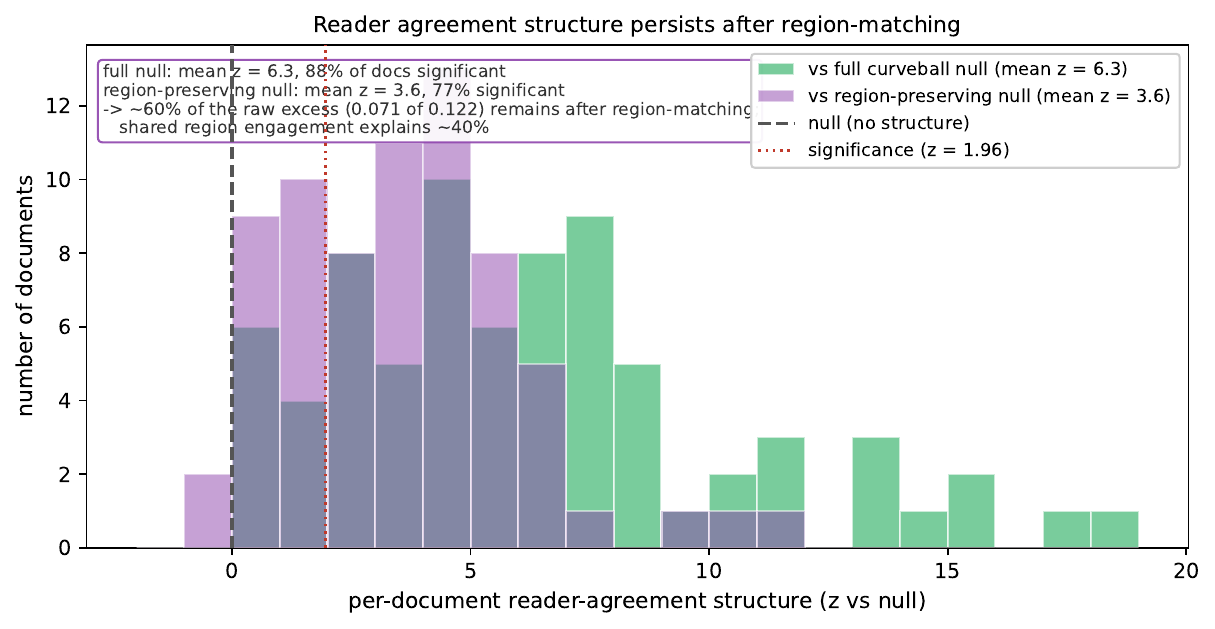}
\caption{Per-document reader-agreement structure (Experiment~1). Each document contributes a
$z$-score for how much its readers cluster beyond a null. Against the full curveball null
(green) the distribution sits far to the right (mean $z=6.3$, $88\%$ significant); against the
stricter region-preserving null (purple) it attenuates but remains positive (mean $z=3.6$,
$77\%$). In agreement units, $\sim$60\% of the excess survives region-matching as a finer
reader-specific component; shared region engagement accounts for only $\sim$40\%.}
\label{fig:structure}
\end{figure}

\paragraph{Structure exists, and it is strong.}
Within-document highlighting has strong, pervasive reader sub-group structure. Across $75$ usable
dense documents (median $25$ readers, $74$ sentences), a reader's best-matching partner agrees
with them at binary cosine $0.588$ observed versus $0.466$ expected under the margin-preserving
null --- an excess of $+0.122$ --- so readers are not independent given shared salience; they
cluster. As a standardized effect this nearest-neighbour agreement is $z=\mathbf{+6.3}$ (95\% CI
$[5.4, 7.3]$, every bootstrap positive), with $\mathbf{66/75=88\%}$ of documents individually
significant ($p\leq 0.05$, against a $5\%$ chance rate); the variance statistic gives $z=+12.8$
($71/75=95\%$ significant). The standardized effect is density-dependent --- mean $z$ ranges from
$3.3$ at median $11$ readers to $6.3$ at median $25$ across samples --- so we report the raw
excess alongside $z$ and use a $\geq 10$-reader sample; the raw decomposition below is stable
across samples. Figure~\ref{fig:structure} (green) shows the distribution far to the right of the
null.

\paragraph{Region engagement is the minority; the majority is finer.}
Is this just readers engaging the same sections? We re-test against the region-preserving null,
which holds each reader's per-region engagement fixed. In agreement units the raw excess falls
from $+0.122$ to $+0.071$ under the region-preserving null, so shared region engagement accounts
for $\sim$$40\%$ of it ($0.051$ of $0.122$) and a finer reader-specific component for the
majority $\sim$$60\%$ ($0.071$ of $0.122$, the part that survives region-matching); this split is
stable across samples (the eight position blocks only approximate true sections, so the surviving
finer share is an upper estimate; see Limitations). As a standardized effect the surviving residual is mean $z=\mathbf{+3.6}$,
with $\mathbf{54/70=77\%}$ of documents still significant (Figure~\ref{fig:structure}, purple). So
part of the within-document structure \emph{is} shared region engagement (readers concentrating on
the same coarse regions), exactly as a critic would expect; but the larger part is a finer
reader-specific agreement --- two readers marking the same \emph{sentences within a region} more
than chance --- that survives even after matching region engagement. In this descriptive sense the
within-document crowd is factional, not merely regionally focused.

\begin{table}[t]
\centering
\small
\begin{tabular}{lccc}
\toprule
 & mean $z$ vs.\ null & docs significant & note \\
\midrule
nearest-neighbour agreement (full null)  & $+6.3$ $[5.4, 7.3]$ & $66/75$ ($88\%$) & every boot.\ $>0$ \\
pairwise-agreement variance (full null)  & $+12.8$ $[10.7, 15.1]$ & $71/75$ ($95\%$) & \\
nearest-neighbour, \emph{region} null    & $+3.6$ & $54/70$ ($77\%$) & finer than regions \\
\midrule
synthetic planted groups (pos.\ control) & $+1.7$ & $50\%$ & test is powered \\
synthetic null (neg.\ control)           & $\approx 0$ & $0\%$ & no false positives \\
\bottomrule
\end{tabular}
\caption{Experiment~1: within-document reader sub-group structure. The full-null effect is
strong and pervasive; shared region engagement explains $\sim$40\% of the raw excess, and the
majority ($\sim$60\%, mean $z=3.6$) survives the region-preserving null as a finer component. The
negative control yields zero false positives, so the signal is not a null artifact; the planted
control confirms the statistic responds in the expected direction (its magnitude is not directly
comparable to real data, as the synthetic parameters are intentionally simplified).}
\label{tab:e1}
\end{table}

\paragraph{Robustness.}
(i) \emph{Calibration:} the no-structure control gives $z\approx 0$ with $0\%$ of documents
significant (no false positives); a planted two-group control gives $z=1.7$, confirming the
statistic responds (Table~\ref{tab:e1}). (ii) \emph{Not a short-document artifact:} the signal
\emph{rises} with document length (mean $z$ of $3.8$, $6.1$, $8.9$ for short/medium/long;
correlation $+0.47$) --- longer documents show \emph{more} structure (and more power: $z$ also
rises with reader count, correlation $+0.52$). (iii) \emph{Not a duplicate-reader artifact:} no
document is driven by near-identical readers ($0/75$ with best-partner cosine $>0.9$).
(iv) \emph{Within-reader anchoring:} two readers ``agree'' only by marking the same sentence, so
there is no cross-reader inflation through sentence similarity.

\paragraph{What we do, and do not, claim here.}
The statistics establish \emph{structured residual agreement} among readers beyond both margins
and (in part) beyond region engagement. They do not, by themselves, recover or validate discrete
``communities''; nearest-neighbour agreement and variance respond to graded heterogeneity as
well as to clean clusters. We use ``sub-group'' and ``faction'' for the document-local
grouping in this descriptive sense, and reserve the question of whether the grouping is a stable,
portable reader \emph{type} for Experiment~2.

\section{Experiment 2: is the structure a stable reader trait?}
\label{sec:e2}
\paragraph{Power first.}
Strong sub-structure on each document does not imply \emph{stable} sub-populations: the groups
could re-form differently every time. We test this with the cross-document split-half
correlation of per-pair excess agreement (\S3), and the decisive issue is power, so we calibrate
first. On synthetic \emph{stable} communities the split-half is large when data is dense ($0.43$
at $24$ readers, $9$ marks, $8$ shared documents), but it \textbf{attenuates with sparsity}
($0.15$ at a realistic $10$ readers / $5$ marks / $3$ shared documents) and \textbf{collapses to
zero power} when pairs co-read only two documents ($-0.08$, indistinguishable from the
\emph{situational} control's $+0.05$). So the test reliably detects only \emph{strong} stable
structure; crucially, pairs that share exactly two documents carry essentially no power.

\paragraph{The same kind of structure is present.}
The Experiment~2 documents also show positive within-document structure (median per-document
$z\approx 3.1$ against the curveball null), at a lower standardized magnitude than Experiment~1
because the stability sample is sparser; the two halves of the paper look at the same phenomenon,
so the stability question is well-posed.

\begin{figure}[t]
\centering
\includegraphics[width=0.84\linewidth]{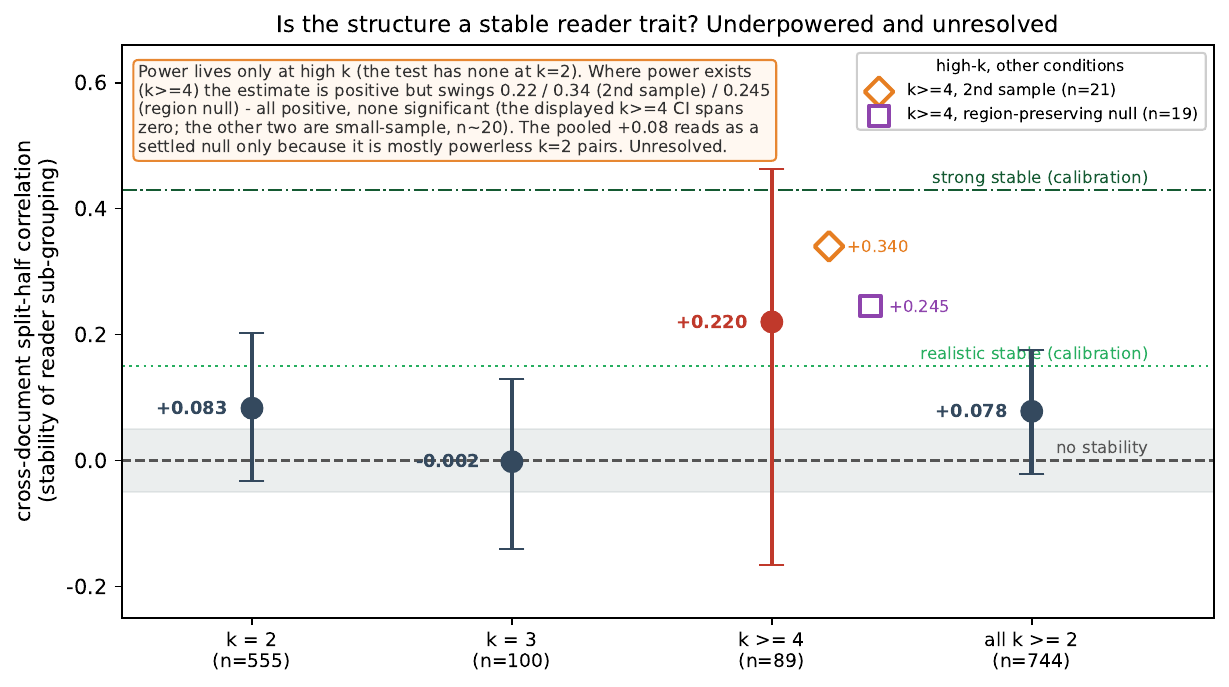}
\caption{Cross-document stability by shared-document count (Experiment~2), against
calibration anchors (no stability $0$; realistic-stable $0.15$; strong-stable $0.43$). The test
has power only at high $k$. Where power exists ($k\geq 4$) the estimate is positive but swings
across conditions ($+0.22$ sample~1, $+0.34$ sample~2, $+0.245$ region-preserving null), none
significant; the displayed $k\geq 4$ interval (sample~1) spans zero, and the other two are
small-sample checks ($n\!\approx\!20$). The pooled estimate averages the only informative subset
into zero-power pairs, so it reads as a settled null when the question is actually undecided.}
\label{fig:calibration}
\end{figure}

\paragraph{Result: unresolved.}
Pooled over all $744$ reader pairs that co-read $\geq 2$ dense documents, the cross-document
split-half is $+0.078$ (95\% CI by pair $[-0.022, 0.175]$; by reader $[-0.057, 0.187]$, so
dyadic dependence is not the issue). The pooled number mixes powered and powerless pairs, so we
stratify by shared-document count $k$ (Table~\ref{tab:e2}, Figure~\ref{fig:calibration}):
$k{=}2$ gives $+0.083$ $[-0.03, 0.20]$ (no power), $k{=}3$ gives $-0.002$ $[-0.14, 0.13]$, and
the informative $k\geq 4$ subset gives $+0.220$ $[-0.17, 0.46]$. The $k\geq 4$ point estimate is
\emph{compatible} with the realistic-stable calibration ($0.15$), but the interval is far too
wide to support that reading, the pattern is non-monotonic in $k$, and \textbf{no stratum is
statistically significant.} It is also imprecise across samples: a separately drawn sample ($414$
pairs) gave a pooled split-half of $0.000$ $[-0.12, 0.13]$ and a $k\geq 4$ estimate of $+0.34$
($n{=}21$, $[-0.39, 0.67]$). The sign is positive in both samples, but every interval includes
zero and the magnitude swings roughly two-fold ($+0.22$ vs.\ $+0.34$) on tiny subsets (all CIs in
Table~\ref{tab:e2}), so neither the existence nor the size of a stable component is established ---
a pattern equally consistent with pure noise and with an under-detected weak effect (the matching
sign \emph{very} weakly favours the latter, but with possible sample overlap is not an independent
confirmation). \textbf{Region check.} Recomputing the excess against the
\emph{region-preserving} null (so any portable signal must be finer than shared region or topic
engagement) attenuates the $k\geq 4$ estimate ($+0.34\!\rightarrow\!+0.245$, $n{=}19$) and leaves
the pooled value at $+0.011$ $[-0.09, 0.11]$ --- consistent with part of the apparent stability being region or
reading-style overlap, which prior work places in the \emph{selection}
layer~\cite{nakayashiki2026selection} rather than within-document salience; but this too is
underpowered. We therefore leave cross-document stability \textbf{unresolved}: the data is
consistent with anything from situational grouping to a weak-to-moderate stable reader trait, and
cannot establish the high-$k$ estimate's sign as reliably non-zero.

\begin{table}[t]
\centering
\small
\begin{tabular}{lccc}
\toprule
shared documents $k$ & pairs & split-half & 95\% CI \\
\midrule
\multicolumn{4}{l}{\emph{Sample 1 ($744$ pairs)}} \\
$k = 2$ (no power)            & 555 & $+0.083$ & $[-0.03, 0.20]$ \\
$k = 3$                       & 100 & $-0.002$ & $[-0.14, 0.13]$ \\
$k \geq 4$ (informative)      & 89  & $+0.220$ & $[-0.17, 0.46]$ \\
pooled $k \geq 2$            & 744 & $+0.078$ & $[-0.02, 0.18]$ (pair); $[-0.06, 0.19]$ (reader) \\
\midrule
\multicolumn{4}{l}{\emph{Robustness: separately drawn sample 2 ($414$ pairs)}} \\
pooled $k \geq 2$             & 414 & $0.000$  & $[-0.12, 0.13]$ \\
$k \geq 4$                    & 21  & $+0.340$ & $[-0.39, 0.67]$ \\
$k \geq 4$, region null       & 19  & $+0.245$ & $[-0.27, 0.54]$ \\
pooled $k \geq 2$, region null & 414 & $+0.011$ & $[-0.09, 0.11]$ \\
\midrule
\emph{calibration: realistic-stable} & --- & $+0.15$ & synthetic ground truth \\
\emph{calibration: strong-stable (dense)} & --- & $+0.43$ & synthetic ground truth \\
\bottomrule
\end{tabular}
\caption{Experiment~2: cross-document stability by shared-document count, with separately drawn
sample-2 robustness rows and synthetic calibration anchors. \emph{No row is significant; every CI
spans zero.} The high-$k$ point estimate is near the realistic-stable anchor but its CI spans zero,
is imprecise across the two samples ($+0.220$ vs.\ $+0.340$), and attenuates under the
region-preserving null ($+0.340\!\rightarrow\!+0.245$). The pooled estimate is diluted by
zero-power $k{=}2$ pairs. Stability is unresolved.}
\label{tab:e2}
\end{table}

\paragraph{Why we cannot resolve it (and why pooling misleads).}
This is a power limit, not a sample-size limit that more crawling fixes cheaply. Resolving a
weak stable component requires pairs that co-read \emph{many} dense documents ($k\geq 4$--$5$),
so a pair's agreement can be averaged into a reliable estimate. Co-readership density (and the
$\sim$50\% article-fetch loss) does not supply such pairs: most pairs that co-read at all share
exactly two fetchable dense documents, where the split-half has no power. Scaling the crawl
raises the \emph{count} of low-$k$ pairs without raising their \emph{information}. The pooling
pitfall is worth stating: a single pooled split-half ($+0.078$ in one sample, $0.000$ in another)
cannot distinguish a confident zero from a zero produced mostly by powerless pairs, so it would
read as a settled null. Stratifying by $k$ shows the pooled value is the latter: the informative
high-$k$ subset is non-zero but too imprecise to settle the question either way. The question is
open, not settled.

\section{Synthesis}
\label{sec:synthesis}
\begin{figure}[t]
\centering
\includegraphics[width=0.88\linewidth]{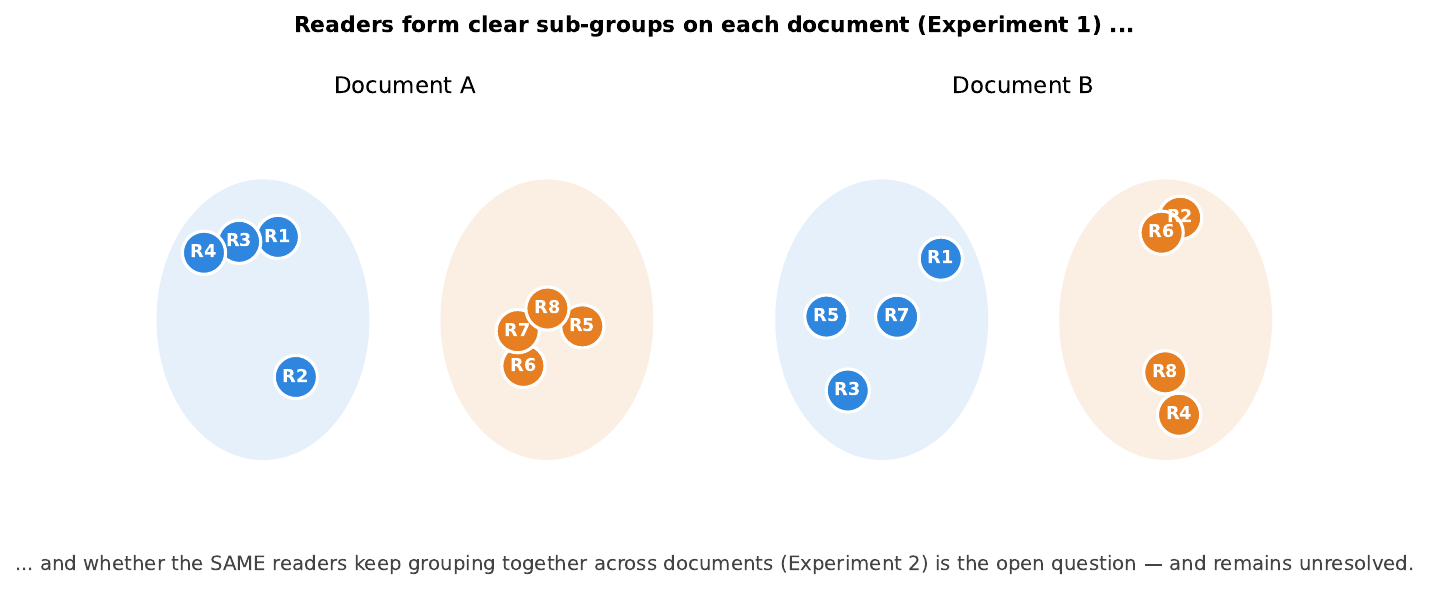}
\caption{The picture. On each document readers form clear sub-groups (Experiment~1); whether the
same readers keep grouping together across documents (Experiment~2) is the open question, and
remains unresolved at the available co-readership density.}
\label{fig:factions}
\end{figure}

The results across reading altitudes are these (Figure~\ref{fig:factions}). Within a document,
the \emph{individual} signal is a whisper (own-versus-other gap
$+0.017$~\cite{nakayashiki2026salience}); the \emph{group} signal in the same place is loud
(readers cluster at $z\approx 6$, only about $40\%$ of it shared region engagement, the majority a
finer residual surviving, \S\ref{sec:e1}); and whether that group structure is a stable, portable reader trait
is, honestly, beyond our reach (\S\ref{sec:e2}). The first two are a clean asymmetry --- a quiet individual inside a
strongly factional crowd. The third is an open question that the available co-readership density
cannot answer, and our best read is a caution against assuming either extreme: the data neither
shows clean situational reshuffling nor confirms stable reader types.

This consistency check should not be over-read as three independent confirmations of one story:
Paper~I's own-versus-other gap and our Experiment~2 are close quantities (both ask whether
within-document marking is a portable reader property, at individual and pair resolution), and
the within-document/region decomposition and the cross-document test share data and machinery.
The honest summary is narrow: the within-document crowd is factional; the portability of its
factions is undetermined.

\section{Discussion}
\paragraph{For products and personalization.}
The strong within-document sub-structure (Experiment~1) is real and could be used \emph{within} a
document --- e.g.\ surfacing two distinct ``ways readers marked this'' rather than one
popular-highlights map, for a document that already has readers. What is \emph{not} supported is
a \emph{portable} segment: we have no evidence that a reader's sub-group can be learned from
their history and applied to a new document, and we cannot even bound how strong such a stable
segment is. It is worth separating two senses of ``cold.'' For a document \emph{no one} has read,
behaviorally grounded reader segments cannot be inferred from that document alone. For a document the \emph{target user} has not read but others have, an
\emph{interaction-conditioned} approach remains open: build the document's sub-groups from its
existing readers and assign a new user after they make their first few marks. Our results bear on
\emph{history-only} segment prediction (unresolved), not on this interaction-conditioned
route, for which Experiment~1's strong per-document structure is a precondition (the accuracy of
assigning a new reader from a few marks is itself untested here).

\paragraph{For digital twins and aggregation.}
A recurring hope is that aggregating many behaviourally-grounded per-user models reconstructs, or
decomposes, collective judgement. Our results temper this for within-document salience: the raw
material (per-document sub-structure) is strong, but the \emph{stability} a history-conditioned
model would have to capture is unconfirmed and not reliably estimable at the current
co-readership density. Where individuality is recoverable
is selection (which documents and topics a person attends to)~\cite{nakayashiki2026selection},
not the within-document salience layer studied here. A faithful ``twin'' of reading is more a
model of \emph{what} a person attends to than of \emph{how} they mark a given page.

\section{Limitations}
\textbf{RQ2 is unresolved, not answered.} Our central caution: Experiment~2 is underpowered for
realistic-strength stability; the informative subset is small ($n{=}89$) with a CI that admits
both zero and a moderate effect. We neither confirm nor exclude a weak stable reader trait.
\textbf{Observable, not latent.} The split-half bounds \emph{observable cross-document
reproducibility} under our co-reading sparsity, not a measurement-error-corrected latent
stability; attenuation (the curveball excess on sparse documents is noisy) biases it downward.
\textbf{High-$k$ confound.} Pairs that co-read many documents may be concentrated in similar
topics, sources, or genres, so any apparent stability in the $k\geq 4$ subset need not imply a
trait portable across arbitrary documents --- and such topical co-selection is exactly where
prior work locates individuality~\cite{nakayashiki2026selection}; the region-preserving null
(\S\ref{sec:e2}) partly probes this and attenuates the estimate.
\textbf{Graded vs.\ discrete.} Experiment~1's statistics detect structured agreement, not validated
discrete communities; we do not claim a specific number of factions, and the two-group cartoon in
Figure~\ref{fig:factions} is illustrative, not a fitted model. \textbf{Region null granularity.}
The region-preserving null uses fixed position blocks as a proxy for sections; true paragraph or
section boundaries would be a sharper control, and because position blocks only partly capture
sections, the surviving finer signal ($z\approx 3.6$) is an \emph{upper} estimate. \textbf{Scope.} The population is doubly
self-selected (people who install a highlighter; documents popular enough to be co-read),
documents skew short and popular (median $\sim$72--83 sentences), and fetch recovers only
$\sim$25--35\% of dense documents (and the current page, which may differ from what a reader saw).
\textbf{One layer.} We study within-document salience only; sub-populations might be stable at
coarser altitudes (topic, source, format), where selection-level
work~\cite{nakayashiki2026selection} locates individuality.

\section*{Ethics and data}
The experiments use aggregate behavioural traces from a social-highlighting platform. We report
only aggregate statistics; per-user and per-pair data are not released because they can reveal
private reading behaviour. The analysis is read-only; profiles are anonymized; only the binary
fact of which sentence a reader marked is used. We do not release copyrighted full text: article
bodies are fetched only transiently for sentence segmentation and are not redistributed, and
Kindle, PDF, and private uploads are excluded by construction.

\section{Conclusion}
When many people highlight the same document, the crowd is not a single consensus: readers form
strong sub-groups that mark systematically different things --- a loud group-level signal exactly
where the individual signal is a whisper --- and about $60\%$ of this is finer than readers merely
engaging the same regions. Whether those sub-groups follow the reader across documents, however,
we cannot determine: the test is underpowered at the available co-readership density, the
informative high-overlap subset is positive but imprecise and never significant
(so neither the existence nor the size of a stable component can be established), and a pooled
estimate would have read as a settled null when the question is actually undecided. The within-document crowd is factional; the portability of its factions is, honestly,
an open question --- and one that limits, for now, what we can responsibly claim about how much
stable reader structure a personalizer can recover.

\appendix
\section{Reproducibility details}
\label{app:repro}
\begin{itemize}
\item \textbf{Re-anchoring.} Article bodies are fetched, segmented into sentences, and each
reader's normalized spans re-anchored to sentences by substring or token-Jaccard ($\geq 0.6$)
match; a reader is kept only if $\geq 50\%$ of their spans re-anchor (drift gate), and dropped if
their anchored marks exceed half the document's sentences (over-anchoring guard).
\item \textbf{Curveball null.} Both margins preserved via trial swaps; per document the swap count
is $\sim$$30\times$ the number of $1$s, well past the mixing point. We use $100$--$150$
permutations; the per-document $z$ is a ratio to the permutation standard deviation rather than a
tail count, so it is not sensitive to the exact permutation count above the mixing point; for the
full-null analysis the $p\leq 0.05$ counts are likewise dominated by documents far from the
threshold (median $z\approx 6$).
\item \textbf{Region-preserving null.} Sentences are split into $8$ equal position blocks as a
proxy for sections; curveball is run independently within each block, preserving each reader's
per-block mark count. The block count trades section fidelity (finer blocks remove more region
structure) against per-block permutation freedom (short documents leave little room to permute
within small blocks); $8$ blocks at a median $\sim$$72$ sentences is $\sim$$9$ sentences per block.
\item \textbf{Experiment~1.} Dense documents have $\geq 10$ co-readers. Statistics: nearest-neighbour
agreement (mean over readers of each reader's max pairwise binary cosine) and the variance of
pairwise cosine; per-document $z=(\text{obs}-\text{mean null})/\text{sd null}$ and permutation $p$;
aggregated with a by-document cluster bootstrap ($3{,}000$ iterations).
\item \textbf{Experiment~2.} Communities sampled from seed documents' co-readers and their history;
dense documents have $\geq 6$ sampled readers. The per-pair excess $e_{AB}(D)$ uses the curveball
(or region-preserving) null mean per document; the split-half resamples a pair's documents into
halves ($200$ random splits), correlated across pairs, with pair-cluster and reader-cluster
bootstraps. Shared-document distribution (sample~1, $744$ pairs): $k{=}2$/$3$/$\geq\!4 =
555$/$100$/$89$; second sample ($414$ pairs): $343$/$50$/$21$. The two samples use the same
parameters but different random seeds (same platform), so readers and documents may overlap ---
they are separately drawn, not statistically independent.
\item \textbf{Calibration.} Synthetic communities with two planted groups, stable vs.\ reshuffled
across documents, at dense ($R{=}24$, $9$ marks, $8$ documents), realistic ($R{=}10$, $5$ marks,
$3$ documents), and sparse ($R{=}8$, $4$ marks, $2$ documents) regimes; the stable/unstable
split-half values anchor Table~\ref{tab:e2}.
\end{itemize}


\begin{thebibliography}{9}
\setlength{\itemsep}{0pt}
\setlength{\parsep}{0pt}
\bibitem{nakayashiki2026salience} K.~Nakayashiki and K.~Watanabe. Personal Salience:
Highlighting Is Social, but Individuality Lives in Selection. arXiv:2606.09024, 2026.

\bibitem{nakayashiki2026selection} K.~Nakayashiki and K.~Watanabe. Selection, Not Salience: The
Shape and Limits of Personalization in Social Highlighting. arXiv:2606.10398, 2026.

\bibitem{winchell2020} A.~Winchell et al. Highlights as an Early Predictor of Student
Comprehension and Interests. Cognitive Science, 2020.

\bibitem{aroyo2015} L.~Aroyo and C.~Welty. Truth Is a Lie: Crowd Truth and the Seven Myths of
Human Annotation. AI Magazine, 2015.

\bibitem{plank2022} B.~Plank. The ``Problem'' of Human Label Variation. EMNLP, 2022.
arXiv:2211.02570.

\bibitem{surowiecki2004} J.~Surowiecki. The Wisdom of Crowds. Doubleday, 2004.

\bibitem{schoenegger2024} P.~Schoenegger et al. Wisdom of the Silicon Crowd: LLM Ensemble
Prediction Capabilities Rival Human Crowd Accuracy. Science Advances, 2024.

\bibitem{park2024} J.~S. Park et al. Generative Agent Simulations of 1,000 People.
arXiv:2411.10109, 2024.

\bibitem{gygli2018} M.~Gygli and M.~Soleymani. PHD-GIFs: Personalized Highlight Detection for
Automatic GIF Creation. ACM MM, 2018.

\bibitem{salemi2023} A.~Salemi et al. LaMP: When Large Language Models Meet Personalization.
ACL, 2024.
\end{thebibliography}
\end{document}